\documentstyle[floats,aps,twocolumn]{revtex}

\begin{document}

\title{Fusion reactions in molecules via nuclear threshold resonances}

\author{V.B. Belyaev, A.K. Motovilov}
\address{Joint Institute for Nuclear Research, Dubna, 141980, Russia}

\author{W. Sandhas}
\address{Physikalisches Institut, Universit\"{a}t Bonn, D-53115 Bonn,
Germany\\[5mm]
\underline{\normalsize{\rm LANL E-print {\tt nucl-th/9508017}; 
Published in J.~Phys.~G {\bf 22} (1996) 1111--1113}}
}

\maketitle
\vspace{1cm}

\begin{abstract}
It is widely accepted that in molecular systems the nuclear interaction
plays a negligible role, because of the strong Coulomb repulsion of the
nuclei at small distances. We are going to show that this is not always
true. The existence of an extended nuclear resonance may lead to
considerably enhanced nuclear reaction rates in appropriately prepared
molecules. Especially we point out that $p + p + ^{16}\!O$, i.e.,
the constituents of water, can form a $^{18}\!Ne (1^-)$ threshold
resonance which decays under energy release into $^{17}\!F$ and a proton.
\end{abstract}

\vspace{1cm}

\pacs{PACS numbers: 24.30.--v, 25.45.--z}

\vspace{1.0cm}
In this note we consider the increase of the reaction 
probability of nuclei in molecules due to the existence of 
nuclear resonances.  Appropriate candidates for such a mechanism 
are molecules in which the nuclei involved can form 
near-threshold resonances. The long tail of the corresponding 
wave functions is expected to lead to a noticeable overlap with 
the molecular functions, and hence to a measurable transition 
probability from the molecular to the nuclear states. For a 
two-atomic molecule it will be shown that an even exponential 
enhancement arises, instead of the usual reduction of nuclear 
cross sections due to Coulomb repulsion.

Let us discuss some examples of such threshold states. A typical 
case is the Boron isotope $^8\!B$.  In its ground state the 
proton has a separation energy of only 0.13~MeV, being thus 
represented by a wave function with a long-ranged tail.  The 
treatment \cite{kim} of the process \, $p+{^7\!Be} \, 
\rightarrow \, ^8\!B + \gamma$, which is crucial in the solar 
neutrino problem~\cite{bel}, did in fact require integrations up 
to 300 fm.

Another example is the $^5\!{He}\,({3/2}^+)$ resonance which 
plays a decisive role in muon catalyzed fusion of deuteron and 
triton in the $dt\mu$ molecule~\cite{muCF}.  The presence of 
this near-threshold resonance (50~keV resonance energy, 
60--70~keV width) enlarges the fusion probability at least by 
four orders of magnitude, as compared with nuclear reactions in 
molecules like $pd\mu$ or $dd\mu$ where no such resonances 
occur.

To achieve a similar enhancement of the fusion probability 
already in normal (electronic) molecules, we have to look for a 
nuclear resonance which lie much closer to the corresponding 
threshold energy. The exited $1^-$ state of the Neon isotope 
$^{18}Ne$ satisfies this requirement. Its experimental energy 
\cite{nero} of 4.522~MeV coincides up to the given digits with 
the threshold energy of the three-body channel $p+p+{^{16}O}$. 
Vice versa, any $pp{^{16}O}$ system, being rotationally excited 
into a $1^-$ state, is energetically degenerate with this 
$^{18}{Ne}(1^-)$ resonance.  In nature this system occurs in 
form of stable, chemically bound water molecules. These 
molecules, being rotationally excited, thus, contain always some 
$^{18}Ne(1^-)$ admixture. And this nuclear state can decay not 
only into the original $p+p+{^{16}\!O}$ channel, but also into a 
two-body channel like $^{17}\!F+p$ , with an energy release of 
0.6~MeV. Excited water molecules, thus, have a non-vanishing 
probability of ``burning'' into this final state.

In order to get some feeling for the corresponding transition 
probabilities, we consider as a more simple example the 
two-atomic $D\,^6\!Li$ molecule. There exists also in this case 
an excited nuclear state, the Beryllium resonance $^8\!Be(2^+)$, 
close to the $d + \, ^6\!Li$  threshold\footnote{This resonance 
can decay into various channels, e.g., \, $n + \, ^7\!Be$, \, $p 
+ \, ^7\!Li$, \, $\alpha + \alpha$, \, with noticeable energy 
yield \cite{selove}.}. Indeed, according to \cite{selove} the 
resonance energy is (22.2 + {\it i} 0.8)~MeV, while the 
threshold, i.e., the energy needed to break up the $^8\!Be$ 
ground state into $d + \, ^6\!Li$, lies at 22.2798~MeV. A 
noticeable interference between the nuclear and molecular wave 
functions, hence, is again to be expected.

For an estimate of the corresponding transition probability we 
use the following model. The resonance state $\psi_{\rm res}(r)$ 
is simply chosen as an outgoing Coulomb s-wave,

\begin{equation}
\label{1}
\psi_{\rm res}(r) \enspace = \enspace \frac{1}{N_{\rm res}} 
\enspace \frac{{\rm e}^{i\eta \ln \kappa r}}{r},
\end{equation}
with $\eta = Z_1 \, Z_2 \alpha \, \sqrt{\mu c^2 / 2E}$\, being 
the Sommerfeld parameter. Here, $\alpha$ is the fine structure 
constant and $E$, the relative energy of the outgoing particles, 
$E=\displaystyle\frac{k^2}{2\mu}$.  The function (\ref{1}) is 
assumed to be normalized to unity within the range of the 
nuclear interaction \cite{shulgina}. This choice of $N_{\rm 
res}$ reflects the nuclear origin of the outgoing particles, 
which move outside the nuclear volume exclusively under the 
influence of the repulsive Coulomb potential. The molecular wave 
function $\psi_{\rm mol}(r)$, representing the motion of the 
nuclei $d$ and $^6\!Li$ under the influence of an effective 
attractive potential with strong Coulomb repulsion at the 
origin, is chosen as a product of the regular Coulomb solution 
$F_0(\kappa,r)$ and an exponentiall decreasing function 
associated with size of the molecule,

\begin{equation}
\label{2}
\psi_{\rm mol}(r) \enspace = \enspace \frac{1}{N_{\rm mol}} \;
\frac{F_0(\kappa,r)}{r} \enspace {\rm e}^{-\kappa r}.
\end{equation}
Within our model, the transition amplitude is given by

\begin{equation}
\label{3}
I \enspace = \enspace \int d^3 r \,
\psi_{\rm res}(r) \psi_{\rm mol}(r) \;.
\end{equation}
In order to calculate this overlap integral we use for the 
regular Coulomb function the representation

\begin{equation}
\label{4}
F_0(\kappa,r) \enspace = \enspace C_0(\eta) r
{\rm e}^{-ir}M(1 - i\eta,2,2ir),
\end{equation}
with

\begin{equation}
\label{5}
M(a,b,z) \enspace = \enspace \frac{\Gamma(b)}{\Gamma(a)\Gamma(b-a)}
\int^1_0 \, {\rm e}^{zt} t^{a-1}(1-t)^{b-a-1}dt,
\end{equation}
and $C_0(\eta) = \exp(- \pi \eta/2)\Gamma(1+i\eta)/\Gamma(2)$. 
This integral representation allows us to perform the 
space-integration in (\ref{3}) analytically. That is, the 
contributions of the wave functions to the transition amplitude 
are taken into account exactly from all distances,

\begin{equation}
\label{6}
I \enspace \propto \enspace 4\pi C_0(\eta) \enspace
\frac{\Gamma(2) \Gamma(2 +
i\eta)}{|\Gamma(1 + i\eta)|^2} \enspace \int^1_0 \enspace
\left(\frac{1 - t}{t}\right)^{i\eta}\times 
\end{equation}
$$
\enspace\times \frac{dt}{\lbrack 1-i(2t-1)
\rbrack^{2+i\eta}}.
$$

\noindent
Since $\eta \gg 1$ near the threshold, the remaining integral 
can be evaluated by means of the saddle point method. The saddle 
point, from which the main contribution to the integral 
(\ref{6}) stems, lies at

\begin{equation}
\label{7}
t_1 \enspace = \enspace 1 + \frac{1}{2} \left[ \sqrt{\sqrt{2} + 1}
+ i \sqrt{\sqrt{2} -1}\,\,\right].
\end{equation}

\noindent
The leading term in (\ref{6}), thus, looks like

\begin{equation}
\label{8}
I \enspace \propto \enspace \frac{C_0(\eta)}{\kappa} \enspace
\frac{\Gamma(2+i\eta)}{|\Gamma(1+i\eta)|^2} \enspace
\frac{{\rm e}^{0.614\pi\eta}}
{\sqrt{\eta}} \left[ A + O\left(\frac{1}{\eta}\right) \right],
\end{equation}
with an $\eta$-independent constant $A$. In other words, we 
obtain for the relevant transition amplitude

\begin{equation}
\label{9}
I \enspace \propto \enspace \eta^{3/2} \exp \lbrack (0.614- 
1/2)\pi \eta \rbrack.
\end{equation}
Thus, opposite to the usual Coulomb barrier factor $\exp(- \pi 
\, \eta)$ which for increasing $\eta$ decreases, we have now an 
increasing factor.  Therefore, due to the fact that instead of a 
short-ranged bound state we deal with a resonance state of long 
range, the contributions to the overlap integral originate 
essentially from intermediate and large distance regions.\\

The present report aimed at pointing out the general property 
(\ref{9}) valid in all situations of the type considered here.  
The specific features of any concrete example enter the problem 
via the reduced mass $\mu$ and the charge numbers $Z_1,Z_2$ in 
the Sommerfeld parameter $\eta$, and via the normalization 
factors $N_{\rm res}$ and $N_{\rm mol}$. By adequately choosing 
these factors, fairly realistic values of the transition 
amplitudes (\ref{3}) can be obtained for the present and all 
analogous cases. Multiplying the squares of these amplitudes by 
$\omega = |\epsilon_{\rm mol}|/ \hbar$, with $\epsilon_{\rm 
mol}$ being the binding energy of the molecular state $\psi_{\rm 
mol}$, one ends up with estimates of the corresponding reaction 
rates. All this will be the subject of practical applications of 
the above formalism to special cases, e.g., the $d + \, ^6\!Li$ 
system.\\

As discussed at the beginning, the rotationally excited $(pp \, 
^{16}O)$ state appears as a promising example for fusion 
reactions in normal (electronic) molecules. Qualitatively a 
similar behaviour as in the two-body case is to be expected also 
in this three-body case, a prediction confirmed by recent 
calculations based on estimates similar to the ones described 
above~\cite{BMS}.\\

\acknowledgments
This work was partly supported by the Scientific Division of 
NATO, grant No. 930102.  Two of the authors (V.B.B. and A.K.M.) 
would like to thank the International Science Foundation for 
financial support, grants No.~RFB000 and No.~RFB300.

\end{document}